# Precision photonic band structure calculation of Abrikosov periodic lattice in type-II superconductors

Alireza Kokabi, Hesam Zandi, Sina Khorasani, Mehdi Fardmanesh[*]

*School of Electical Engineering, Sharif University of Technology, P. O. Box 11365-9363, Tehran, Iran*

**Abstract**

We have performed a numerical solution for band structure of an Abrikosov vortex lattice in type-II superconductors forming a periodic array in two dimensions for applications of incorporating the photonic crystals concept into superconducting materials with possibilities for optical electronics. The implemented numerical method is based on the extensive numerical solution of the Ginzburg-Landau equation for calculating the parameters of the two-fluid model and obtaining the band structure from the permittivity, which depends on the above parameters and the frequency. This is while the characteristics of such crystals highly vary with an externally applied static normal magnetic field, leading to nonlinear behavior of the band structure, which also has nonlinear dependence on the temperature. The similar analysis for every arbitrary lattice structure is also possible to be developed by this approach as presented in this work. We also present some examples and discuss the results.

## 1. Introduction

Recently, the concept of photonic band structure in type-II superconductors and its tunability and controllability with external parameters especially magnetic field are investigated [1,4]. In type-II superconductors when the applied external magnetic field is above the first critical field, a lattice of vortices is formed which depend on magnitude and direction of the applied magnetic field. Hence a permittivity contrast is obtained inside and outside of the vortices. This periodic structure is in the form of a periodic distribution of permittivity normal to the superconductor slab, which represents a two-dimensional photonic crystal (2D-PC). So far the estimated calculation of band structure with simplified models for permittivity with step variation consideration has been reported [1]. Recently, an accurate approach to the numerical computation of order parameters in Ginzburg-Landau equations with external magnetic field between the first and second critical magnetic fields is done [2]. By using this approach one can determine the continuous variation of permittivity along the superconductor. Generally the permittivity obtained is frequency-dependent, and thus the standard plane wave expansion (PWE) method would not be applicable to these calculations. Hence we have used the recently-reported revised plane wave expansion (RPWE) [3], which properly takes the effect of the dispersion and losses into account. By combining these two methods we have computed the precise band structure of the Abrikosov lattice. We have also compared these results with previous estimated calculations.

## 2. Theoretical Model and Analysis

By two-fluid model we can find the permittivity with respect to order parameter in all along the superconductor. The effective dielectric constant is here given by [5]

$$\varepsilon_{eff} = \varepsilon\left\{1 - \frac{\omega_{ps}^2(x,y)}{\omega^2} - \frac{\omega_{pn}^2(x,y)}{\omega[\omega + i\gamma(x,y)]}\right\} \quad , \quad (1)$$

where $\omega_{ps}$ and $\omega_{pn}$ indicate the plasma frequency of superconducting and normally conducting electrons, respectively, and $\gamma$ represents the damping term in the normal conducting states; $\varepsilon$ is the dielectric constant of superconductor. In general, $\omega_{ps}$ and $\omega_{pn}$ depend on superconducting $n_s$ and normal electron densities $n_n$. It is obvious that the sum of these two densities is constant throughout, equaling to the total electron density.

[*] Corresponding author. Tel.: +98-21-6616-5920; fax: +98-21-6602-3261; e-mail: fardmanesh@sharif.edu.

For calculating the effective dielectric constant we need to calculate the densities by order parameter $\Psi$ in Ginzburg-Landau equation. Here, one has three nonlinearly coupled equations in terms of the Fourier series coefficients of normalized $|\Psi|^2$, magnetic field $B$, and supervelocity $\mathbf{Q}$.

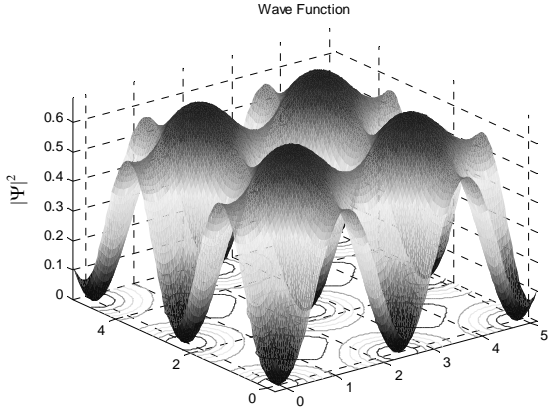

Fig. 1. Normalized superelectron density for $\kappa = 1.5$ and B = 0.7.

For considering the dispersion effect of permittivity we use the revised plane wave expansion method (RPWE) and the Fourier series coefficients of permittivity obtained by (1). In RPWE the normalized frequency $f$ and $\kappa_x$ (wave-vector in $x$-direction) are scanned and $\kappa_y$ is obtained as eigenvalues of a matrix as a result of combining Maxwell's equations. The photonic band gap (PBG) depends on the distances between vortices, and the intensity of applied magnetic field. The resulting periodic perturbations of the permittivity give rise to bandstructure whose PBGs and cut-off frequencies vary with the applied magnetic field.

Solution of Ginzburg-Landau obtained [2] may be applied to square and triangular vortex lattices. Figure 1 shows a typical result for Ginzburg-Landau solution. It shows $|\Psi|^2 = n_s/n$ versus position in the superconductor. By solving the recursive equations, the Fourier expansion coefficient of superconducting electron densities is obtained and then both plasma frequencies as shown in (1) are computed. Having the permittivity known with respect to the position in superconductor, the corresponding Fourier coefficients can be obtained. In our calculation method, the coded algorithm can be equally applied to both standard square and triangular lattice geometries. Then the series coefficients are applied to bandstructure calculations. Damping term, applied magnetic field and Ginzburg-Landau parameter are the three major parameters that affect the band structure. For high dampings above the scanned frequencies, variation of permittivity decrease and thus we have lower cut-off frequencies. Here we have done our analysis for dampings much more than analyzing frequency. Thus the results could be compared with previous estimated ones.

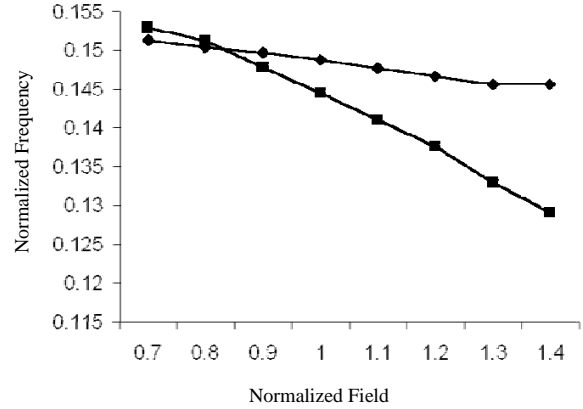

Fig. 2. Estimated (Squares) and precise (Diamonds) results for variations of normalized cut-off frequency versus magnetic field.

As it is apparent from Fig. 2, the cut-off frequency strongly depends on the magnetic field and with increasing the magnetic field the cut-off frequency decreases. It is also concluded that in precise calculation, falling rate for cut-off frequency is lower than that of approximate calculations for a given value of $\gamma$.

## 3. Conclusion

We solved the coupled Ginzburg-Landau and Photonic Band Gap equations, with full accuracy based on two-fluid model. A general code was developed capable of solving for arbitrary geometry and set conditions [6]. We noticed that the cut-off frequency decreases significantly with the applied magnetic field. This effect provides a window to novel potential applications such as precision magneto-optical tunable filters and modulators. The dependence of other parameters of the band structure on the applied field and the vortex lattice is under further investigation.

## 4. References

[1]  H. Takeda, K. Yoshino, Phys. Rev. B **70**, 085109 (2004)
[2]  E.H. Brandt, Phys. Rev. Let. **78**, 2208 (1997)
[3]  Sh. Shi, C. Chen, D. W. Prather, Appl. Phys. Let. **86**, 043104 (2005)
[4]  C. H. R. Ooi, T. C. A. Yeung, C. H. Kam, T. K. Lim, Phys. Rev. B **61**,5922 (2000)
[5]  S. Caorsi, A. Massa, M. Pastorino, IEEE Trans. Microwave Th. Tech. **49**,10 (2001)
[6]  H. Zandi, A. Kokabi, A. Jafarpour, S. Khorasani, M. Fardmanesh, A. Adibi, Proc. SPIE **6480**, to be published.